\documentclass[12pt,epsf]{iopart}
\input epsf
\begin{document}

\title{Beam Simulation Tools for
GEANT4 (and Neutrino Source Applications)}

\author{V.\ Daniel Elvira\dag\ , Paul Lebrun\dag\ , and \\
Panagiotis Spentzouris\dag\ 
\footnote[3]{Correspondence should be addressed to V.\ Daniel Elvira (daniel@fnal.gov)}
}

\address{\dag\ Fermilab, P.O.\ Box 500, Batavia, IL 60555, USA}

\begin{abstract}
Geant4 is a tool kit developed by a collaboration of physicists and 
computer professionals in the High Energy Physics field for simulation of
the passage of particles through matter. The motivation for the development 
of the Beam Tools is to extend the Geant4 applications to
accelerator physics. Although there are many computer programs for beam 
physics simulations, Geant4 is ideal to model a beam going through 
material or a system with a beam line integrated to a complex detector. 
There are many examples in the current international High Energy Physics 
programs, such as studies related to a future Neutrino Factory, a Linear 
Collider, and a very Large Hadron Collider.
\end{abstract}



\maketitle

\section{Introduction}

Geant4 is a tool kit developed by a collaboration of physicists and computer 
professionals in the High Energy Physics (HEP) field for simulation of the 
passage of particles through matter.
The motivation for the development of the Beam Tools is to extend 
the Geant4 applications to accelerator physics. 
The Beam Tools are a set of C++ classes designed to 
facilitate the simulation of
accelerator elements such as 
r.f. cavities, magnets, absorbers. These elements
are constructed from the standard Geant4 solid volumes such as 
boxes, tubes, trapezoids, or spheres. 

A variety of visualization packages are available within the Geant4 
framework to produce an image of the 
simulated apparatus. The pictures shown in this article
were created with Open Inventor~\cite{inventor}, which
allows direct manipulation of the objects on the screen, 
plus perspective rendering via the use of light.

Although there are many 
computer programs for beam physics simulations, Geant4 is ideal
to model a beam through a material or to integrate a
beam line with a complex detector. There are many such examples in the current 
international High Energy Physics programs.

\section{A Brief Introduction to Geant4}

Geant4 is the object oriented C++ version of the Geant3 tool
kit for detector simulation developed at CERN. 
It is currently being used in many fields, such us HEP, space exploration, 
and medicine.

As a tool kit, Geant4 provides a set of libraries, 
a \verb"main" function, and a family of 
initialization and action classes to be implemented by 
the user. These classes are singlets, and their
associated objects are constructed in \verb"main". The objects 
contain the information related to the geometry of the 
apparatus, the fields, the beam, and actions taken by the user at different 
times during the simulation. The Geant4 library classes start with the
\verb"G4" prefix. The example described in this section, called
\verb"MuCool", uses only some of the many available user classes. 

\subsection{Detector and Field Construction}

The detector and field geometry, properties, and location are
implemented in the constructor and methods of the \verb"MuCoolConstruct" user
class, which inherits from \verb"G4VUserDetectorConstruction".
In the \verb"Construct()" method the user does the 
initialization of the electromagnetic field and the equation of motion.
There are a variety of Runge-Kutta steppers to select from, which perform
the integration to different levels of accuracy.
Next comes the detector description, which involves the construction
of solid, logical, and physical volume objects. They contain information
about the detector geometry, properties, and position, respectively. 
Many solid types, or shapes, are available. For example,
cubic (box) or cylindric shapes (tube), are constructed as:

\begin{verbatim}
G4Box(const G4String& pName, G4double pX, G4double pY, G4double pZ);
G4Tubs(const G4String& pName, G4double pRMin, G4double pRMax, 
       G4double pDz, G4double pSPhi, G4double pDPhi);
\end{verbatim}
  
where a name and half side lengths are provided for the box. Inner, outer
radii, half length, and azimuthal coverage are the arguments of a 
cylinder (tube).
A logical volume is constructed from a pointer to a solid, and a given 
material:

\begin{verbatim}
G4LogicalVolume(G4VSolid* pSolid, G4Material* pMaterial,
                const G4String& name)
\end{verbatim}

The physical volume, or placed version of the detector is constructed as:

\begin{verbatim}
G4PVPlacement(G4RotationMatrix *pRot, const G4ThreeVector &tlate, 
              const G4String& pName, G4LogicalVolume *pLogical, 
              G4VPhysicalVolume *pMother, G4bool pMany, G4int pCopyNo);
\end{verbatim}

where the rotation and translation are performed with
respect to the center of its ``mother'' volume (container).
Pointers to the associated logical volume, and the copy number complete the
list of arguments.

\subsection{Physics Processes}

Geant4 allows the user to select among a variety of physics processes 
which may occur during the interaction of the incident particles with the 
material of the simulated apparatus. 
There are electromagnetic, hadronic and other
interactions available like: ``electromagnetic'', ``hadronic'', 
``transportation'', ``decay'', ``optical'', ``photolepton\_hadron'',
``parameterisation''. 
The different types of particles and processes are created in the
constructor and methods of the \verb"MuCoolPhysicsList" user class, which
inherits from \verb"G4VUserPhysicsList".

\subsection{Incident Particles}\label{sec:beam}

The user constructs incident particles, interaction verteces, or a beam 
by typing code in the constructor and methods
of the \verb"MuCoolPrimaryGeneratorAction" user
class, which inherits from \newline
\verb"G4VUserPrimaryGeneratorAction".

\subsection{Stepping Actions}\label{sec:stepping}

The \verb"MuCoolSteppingAction" user action class inherits from
\verb"G4UserSteppingAction". It allows to perform actions
at the end of each step during the integration of the equation of motion.
Actions may include killing a particle under certain conditions,
retrieving information for diagnostics, and others.

\subsection{Tracking Actions}

The \verb"MuCoolTrackingAction" user action class inherits from
\verb"G4UserTrackingAction". For example, particles may be killed here
based on their dynamic or kinematic properties.

\subsection{Event Actions}

The \verb"MuCoolEventAction" user action class inherits from
\verb"G4UserEventAction". It includes
actions performed at the beginning or
the end of an event, that is immediately before or after a particle is
processed through the
simulated apparatus.

\section{Description of the Beam Tools Classes}

This Section is devoted to explain how to simulate accelerator 
elements using the Beam Tools. Brief descriptions of each class and
constructor are included.

\subsection{Solenoids}

The Beam Tools provide a set of classes to simulate realistic solenoids.
These are {\tt BTSheet}, {\tt BTSolenoid}, {\tt BTSolenoidLogicVol}
and {\tt BTSolenoidPhysVol}. 

\begin{itemize}
\item The \verb"BTSheet" class 
inherits from \verb"G4MagneticField". The class objects are
field maps produced by an infinitesimally thin solenoidal 
current sheet.
The class data members are all the parameters necessary to generate
analytically a magnetic field in $r$-$z$ space (there is $\varphi$ symmetry).
No geometric volumes or materials are associated with the \verb"BTSheet" 
objects. \verb"GetFieldValue" is a concrete method of \verb"BTSheet"
inherited from \verb"G4Field", through \verb"G4MagneticField". 
It returns the field value at a given point in space and time.
\item The \verb"BTSolenoid" class inherits from 
\verb"G4MagneticField". The class objects are
field maps in the form of a grid in $r$-$z$ space, which are generated by a 
set of \verb"BTSheet". The sheets and the \verb"BTSpline1D" objects, 
containing the  spline fits of $B_z$ and $B_r$ versus $z$ for each $r$ in the 
field grid, are data members of \verb"BTSolenoid".
No geometric volumes or materials are associated with \verb"BTSolenoid".
The field at a point in space and time is accessed through a 
\verb"GetFieldValue" method, which performs a linear interpolation in $r$ of
the spline fit objects.
\item The \verb"BTSolenoidLogicVol" class defines the material and 
physical size of the coil system which is represented by the set of current 
sheets. A \verb"BTSolenoid" must first be constructed from a 
list of current \verb"BTSheets".  The \verb"BTSolenoid" object is a data
member of \verb"BTSolenoidLogicVol".
The \verb"BTSolenoidLogicVol" class constructor creates 
\verb"G4Tubs" solid volumes and associated logical volumes for 
the coil system, the shielding, and the empty cylindric regions inside them.
Only the logical volumes are constructed here. 
No physical placement of a magnet 
object is done.
\item The \verb"BTSolenoidPhysVol" class is the placed 
version of the \newline
\verb"BTSolenoidLogicVol".
It contains the associated \verb"BTSolenoid" object as a data member, as well
as the pointers to the physical volumes of its logical constituents.
\end{itemize}

\begin{figure}[h]
\hskip-4cm
\vspace{1mm}
\epsfxsize=2.6in
\centerline{\epsfbox{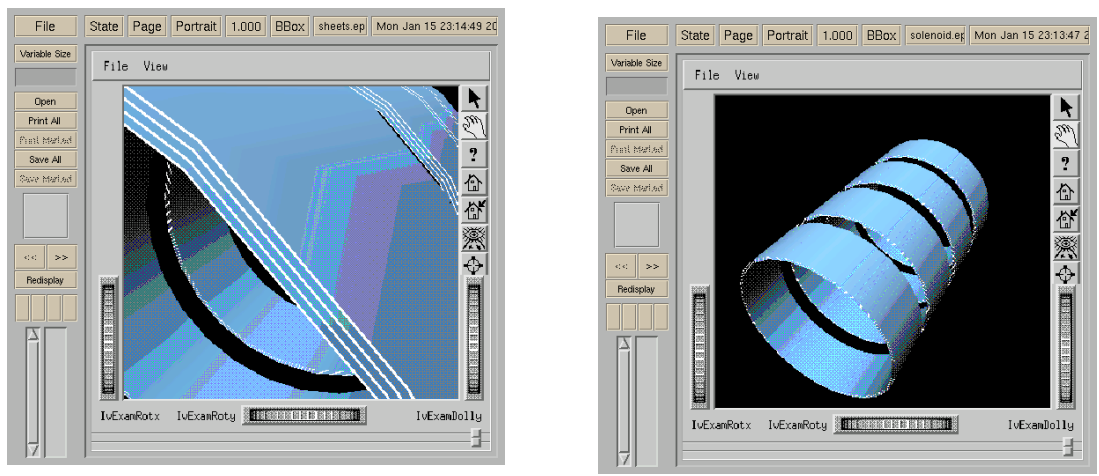}}
\caption{Left: a solenoidal copper coil system 
modeled with four infinitesimally thin sheets equally spaced in radius.
Right: array of four solenoids separated by gaps.}
\label{fig:sheets}
\end{figure}

Figure~\ref{fig:sheets} shows a group of four
solenoidal copper coil systems modeled with four infinitesimally thin sheets
equally spaced in radius. 

\begin{figure}[h]
\vskip-8.6cm
\vspace{1mm}
\hskip4cm
\epsfxsize=2.6in
\centerline{\epsfbox{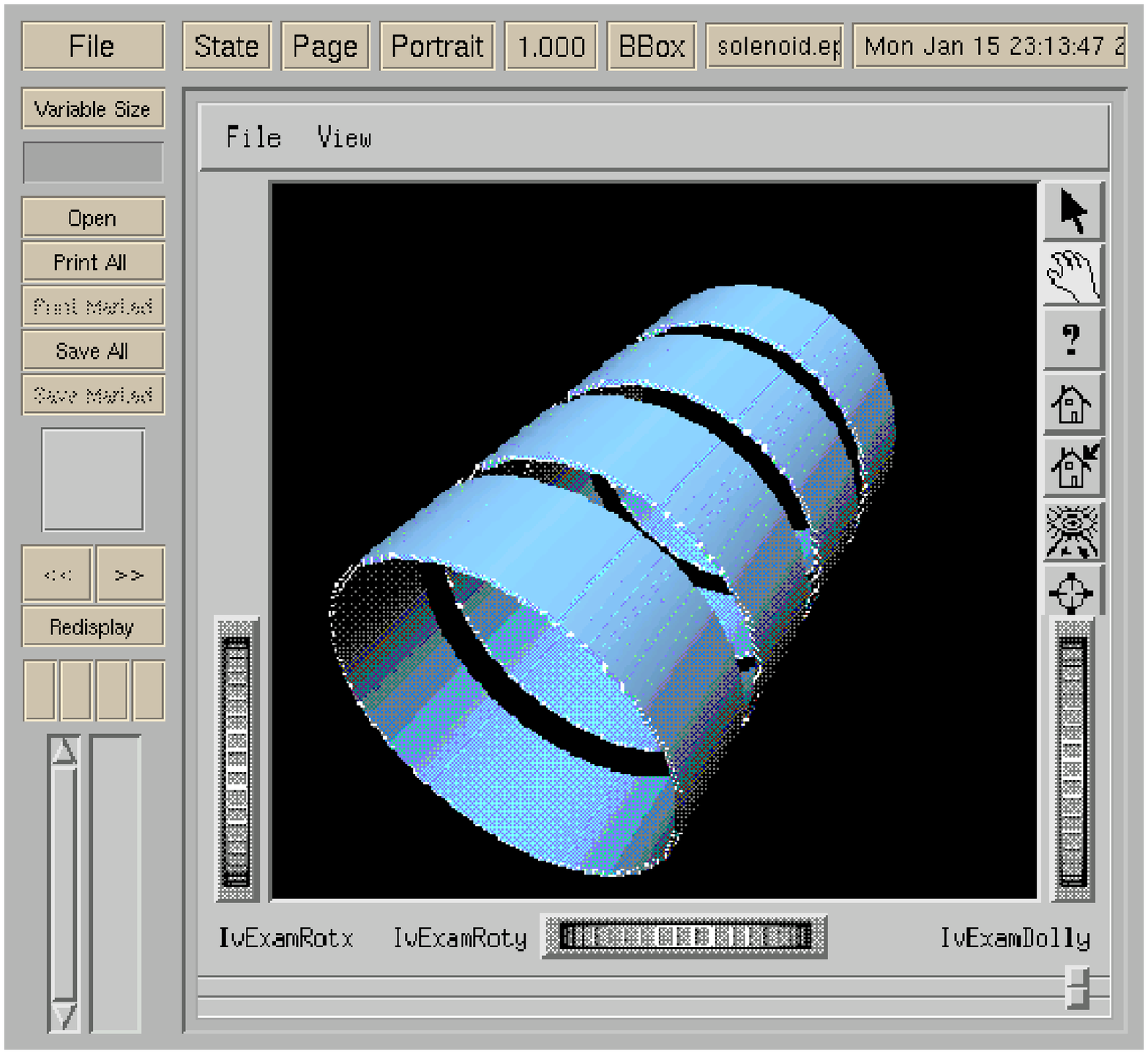}}
\vskip1.2cm
\label{fig:solenoid}
\end{figure}

\subsection{Magnetic Field Maps}

The Beam Tools also allow to simulate generic field maps using the
\verb"BTMagFieldMap" and \verb"BTMagFieldMapPlacement" classes.

\begin{itemize}
\item \verb"BTMagFieldMap" class inherits from
\verb"G4MagneticField". The constructor reads the map information from an 
\verb"ASCII" file containing the value of the field at a set of nodes of a 
grid. No geometric objects are associated with the field. 
The field at a point in space and time is accessed through a 
\verb"GetFieldValue" method, as in the case of the solenoid. 

\item The \verb"BTMagFieldMapPlacement" class is a 
placed \verb"BTMagFieldMap" object. Only the field is placed because there is 
no coil or support system associated with it.
\end{itemize}

\subsection{r.f. Systems: Pill Box Cavities and Field Maps}\label{sec:rfsystems}

This section explains how to simulate realistic r.f. systems using
Pill Box cavities. The Beam Tools package provides the classes: 
\verb"BTAccelDevice", \verb"BTPillBox", \verb"BTrfCavityLogicVol", 
\verb"BTrfWindowLogicVol", and \verb"BTLinacPhysVol". 

\begin{itemize}
\item \verb"BTAccelDevice.hh" class is abstract.
All accelerator device classes are derived from this class, which inherits 
from {\tt G4ElectroMagneticField}.

\item The \verb"BTPillBox" class inherits from 
\verb"BTAccelDevice" and represents single $\pi/2$ Pill Box field objects. 
No solid is associated with \verb"BTPillBox".
The time dependent electric field 
is computed using a simple Bessel function. It is accessed through a 
\verb"GetFieldValue" method.
The field is given by:

\begin{eqnarray}
E_{z} =  V_{p} \, J_{0} \left( \frac{2\pi \nu}{c} r 
\right) {\mathrm sin}(\phi_s + 2\pi\nu t )  \\
B_{\mathrm \varphi} = 
\frac{V_{p}}{c} J_{1} \left( \frac{2\pi \nu}{c} r 
\right) {\mathrm cos}(\phi_s+ 2\pi\nu t )
\label{eq:pillbox1}
\end{eqnarray}

where $V_{p}$ is the cavity peak voltage, $\nu$ the wave
frequency, $\phi_s$ the synchronous phase, and $J_{0,1}$ the Bessel
functions evaluated at $\left( \frac{2\pi \nu}{c} r \right)$. 

\item The \verb"BTrfMap" class also inherits from {\tt BTAccelDevice}. 
The class objects are electromagnetic field maps which represent an r.f. 
cavity. In this way, complex r.f. fields can be measured or generated and 
later included in the simulation. The field map, in the
form of a grid, is read in the \verb"BTrfMap" constructor from 
an \verb"ASCII" file.
The \verb"BTrfMap" object is a field, with no associated solid.
A \verb"GetFieldValue" method retrieves the field value at a point in space and
time.

\item The \verb"BTrfCavityLogicVol" class constructor creates
solid and logical volumes associated with the r.f. field classes.
In the case of a map, a vacuum cylinder ring represents its 
limits. In addition to geometric and material parameters of the cavity, 
the class contains field and accelerator device information. 

\item The \verb"BTrfWindowLogicVol" class  is used with 
\verb"BTCavityLogicVol" to create the geometry and logical volume of 
r.f. cavity windows, including the support structure, which may be placed to
close the cavity iris at the end cups.

\item The \verb"BTLinacPhysVol" class is a placed linac object. 
A linac is a set of contiguous r.f. cavities, including the field, 
the support and conductor material, and windows.
The \verb"BTLinacPhysVol" constructor is overloaded. One version
places a linac of Pill Box cavities and the other places field maps.
\end{itemize}

\begin{figure}[h]
\vspace{1mm}
\hskip-4cm
\epsfxsize=2.6in
\centerline{\epsfbox{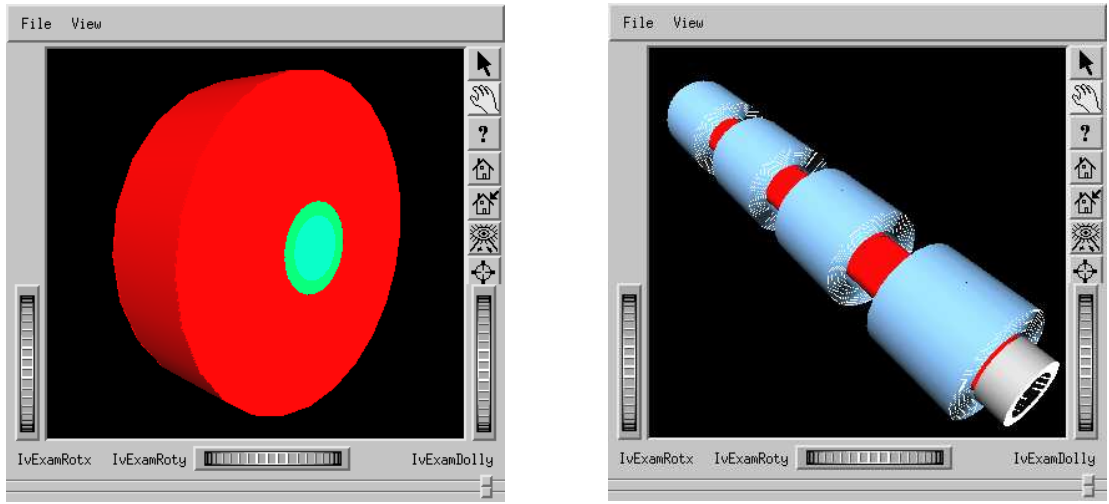}}
\caption{Left: a Pill Box cavity (in red), with windows (green). Right: 
low frequency cooling channel.
The red cylinders are the dummy software structure representing the limits
of the electric field maps.}
\label{fig:berywin}
\end{figure}

Fig.~\ref{fig:berywin} shows a Pill Box cavity (in red) with windows. It
also shows a cooling channel
where solenoids are embedded in large low frequency cavities.
Since the beam circulates inside the solenoid, the cavity is represented
by a field map (in red) restricted to a cylindric volume with radius 
slightly smaller than the inner radii of the magnets.

\begin{figure}[h]
\vskip-9cm
\vspace{1mm}
\hskip4cm
\epsfxsize=2.6in
\centerline{\epsfbox{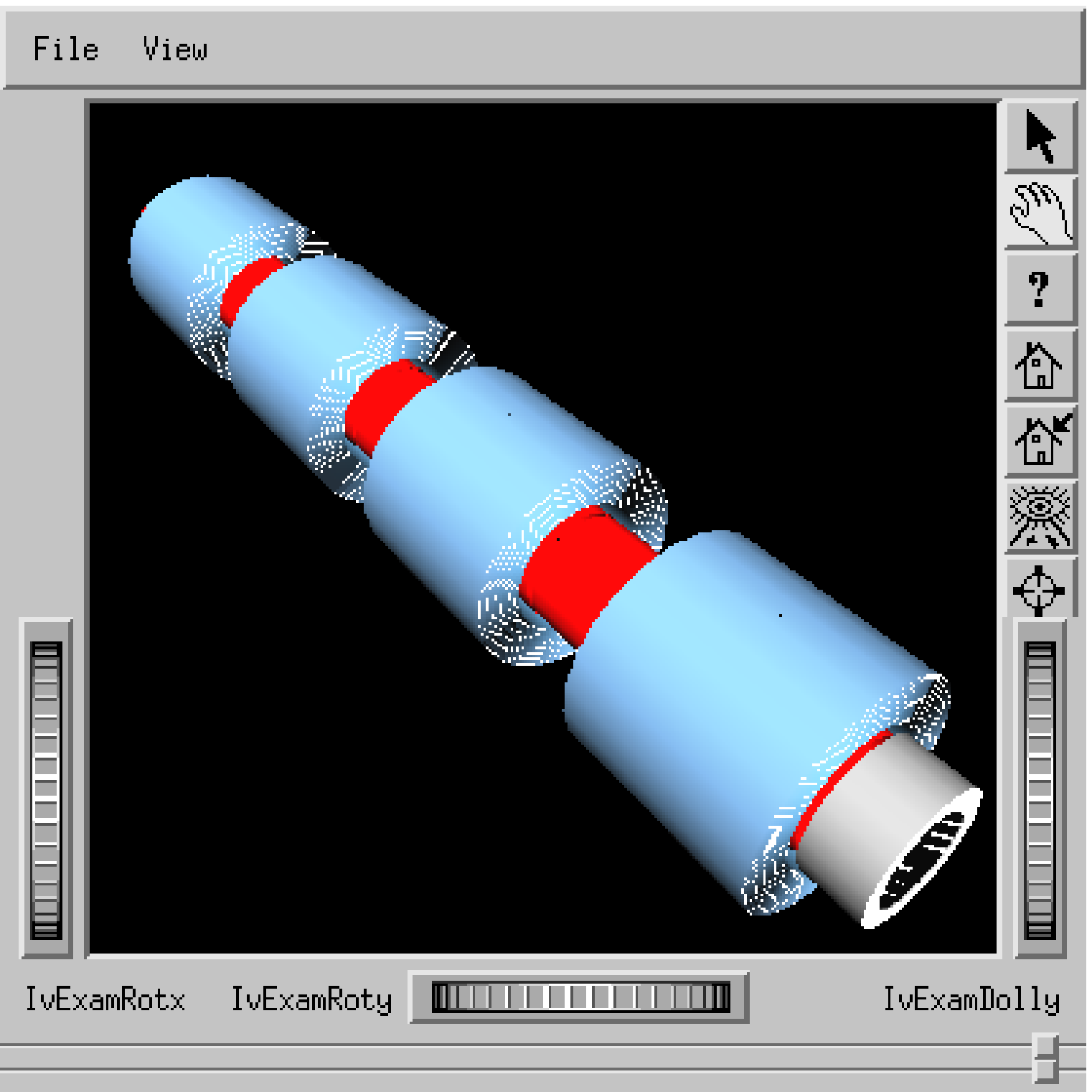}}
\vskip1.5cm
\label{fig:detailview}
\end{figure}

\subsection{Tuning the r.f. Cavity Phases}\label{sec:tune}

One of the critical elements of an accelerator simulation is the ``r.f.
tuning''. Each cavity must be operated at the selected synchronous phase
at an instant coincident with the passage of the beam.
The r.f. wave must be therefore synchronized with the beam, 
more specifically, with the region of beam phase space that the user
needs to manipulate. For this, there is
the concept of a reference particle, defined as the particle with velocity
equal to the phase velocity of the r.f. wave. 
If the kinematic and dynamic variables
of the reference particle are set to values which are 
coincident with the mean values of the corresponding variables for
the beam, the r.f. system should affect the mean beam properties in a similar
way it affects the reference particle. 

The Beam Tools allow the use of a ``reference particle'' to tune the r.f. 
system before processing the beam. The time instants
the particle goes through the phase center of each cavity are calculated
and used to adjust each cavity phase to provide 
the proper kick, at the selected synchronous phase. 

\subsection{Absorbers}

The Beam Tools provide a set of classes to simulate blocks of material
in the path of the beam. The constructors create the solid, logical,
and physical volumes in a single step. They are all derived from the abstract 
class of absorber objects \verb"BTAbsObj".

\begin{itemize}
\item \verb"BTCylindricVessel" is a 
system with a central cylindric rim, and two end cup rims with thin 
windows of radius equal to the inner radius of
the vessel. The material is the same for the vessel walls and windows,
and the window thickness is constant. The vessel is filled with an absorber
material.

\item Two classes are available to simulate absorber 
lenses: \\
\verb"BTParabolicLense" and \verb"BTCylindricLense". The first one is a
class of parabolic objects with uniform density, and the second a cylinder 
object with the density decreasing parabolically as a function of radius. 
From the point of view of the physics effect on the beam,
both objects are almost equivalent.
The \verb"BTParabolicLense" 
is built as a set of short cylinders. The radius is maximum for
the central cylinder and reduces symmetrically following a parabolic equation 
for the others in both sides.
The \verb"BTCylindricLense" object is built from concentric 
cylinder rings of the same length, different radius, and different densities 
to mimic a real lens.
\end{itemize}

\begin{figure}[h]
\vspace{1mm}
\hskip-4cm
\epsfxsize=2.6in
\centerline{\epsfbox{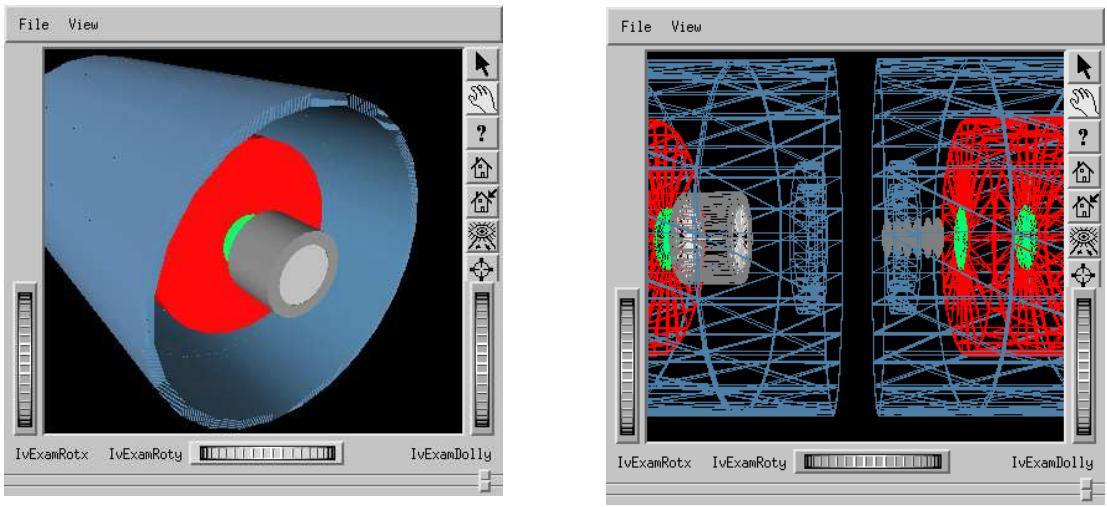}}
\caption{Left: cooling unit cell composed of a solenoid (blue), 
surrounding the r.f. system (red) and the cylindric absorber vessel (gray).
Right: six parabolic lenses (gray) inside a complex magnetic field.}
\label{fig:unitback}
\end{figure}

The gray cylinder in Fig.~\ref{fig:unitback} 
is a schematic representation of a liquid hydrogen vessel
with aluminum walls and windows.
Figure~\ref{fig:unitback} also shows a set of six parabolic lenses in the
center of a complex magnetic system.
The lenses are placed to mitigate the effect
of the decrease in $\langle p_z \rangle$ at large radii in a magnetic field
flip region, using an emittance exchange mechanism.

\begin{figure}[h]
\vskip-9.5cm
\vspace{1mm}
\hskip4cm
\epsfxsize=2.6in
\centerline{\epsfbox{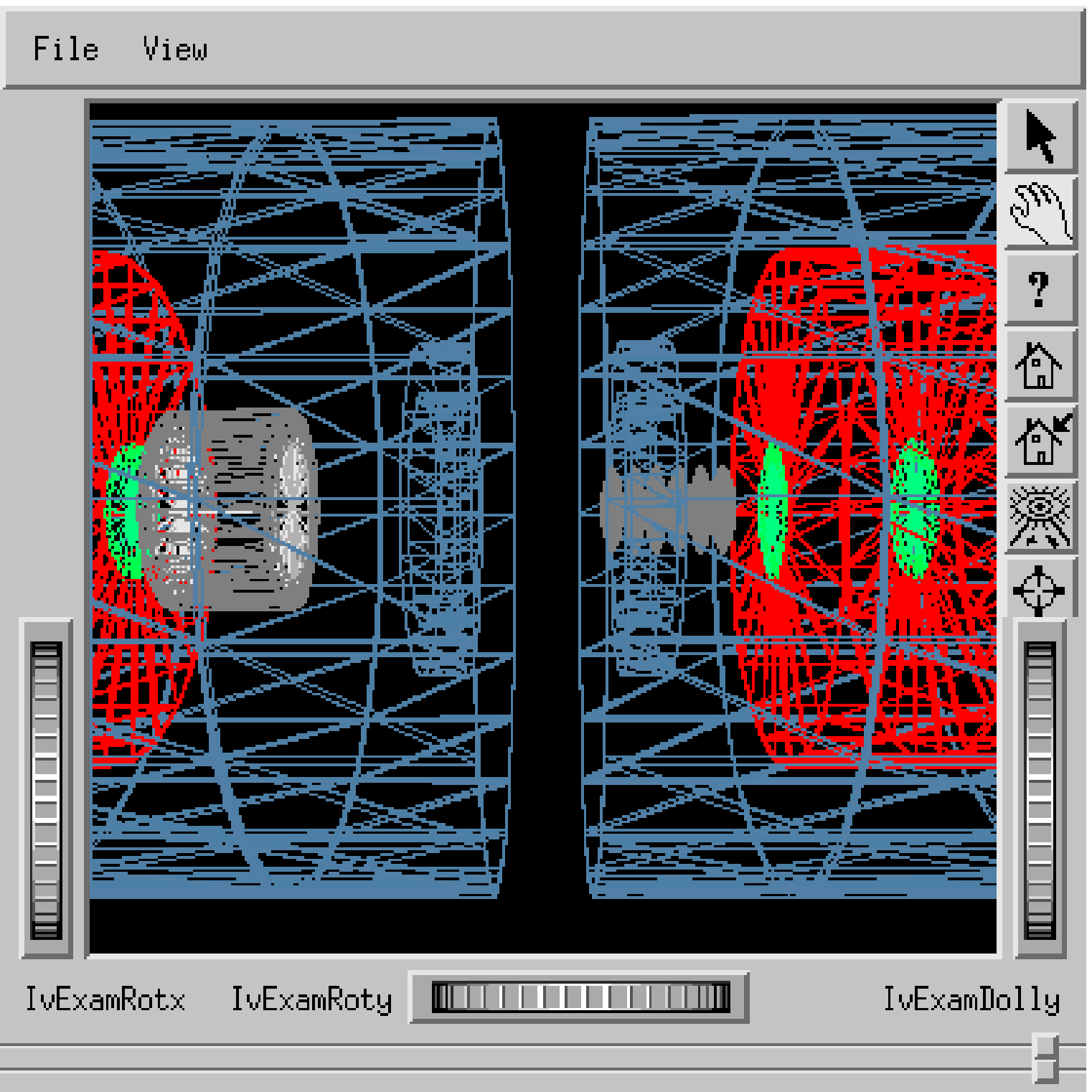}}
\vskip1.5cm
\label{fig:paraflip}
\end{figure}

Wedge absorbers are also useful in some cases. 
They can be easily constructed using the Geant4
trapezoid shape \verb"G4Trap". 

\section{Applications to Neutrino Factory Feasibility Studies}

The neutrino beam in a Neutrino Factory would be the product of the decay
of a low emittance muon beam. Muons would be the result of pion decay, and 
pions would be the product of the interaction of an intense proton beam with 
a carbon or mercury target. Thus the challenge in the design and construction
of a Neutrino Source is the muon cooling section, aimed to reduce the
transverse phase space by a factor of ten, to a transverse emittance of
approximately $\varepsilon_x \sim$1~cm.

The ionization cooling technique uses a combination of linacs and light 
absorbers to reduce the transverse emittance of the beam, while 
keeping the longitudinal motion under control. There are two competing terms
contributing to the change of transverse emittance $\varepsilon_x$ along
the channel. One is a cooling term, associated with the process of 
energy loss, and the other is a heating term related to multiple 
scattering. 

\subsection{The Double Flip Cooling Channel}

\begin{figure}[h]
\vspace{1mm}
\hskip-4cm
\epsfxsize=2.6in
\centerline{\epsfbox{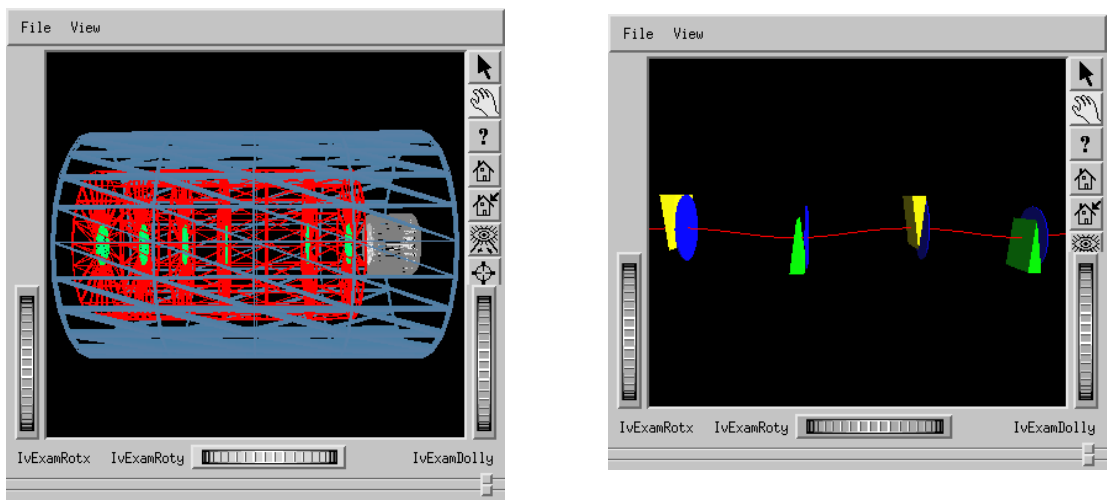}}
\caption{Left: side view of the double flip channel unit cell, including
the solenoid, the six Pill Box cavities, and the absorber.
Right: image of the helical channel, including
the wedge absorbers (yellow and green), idealistic thin r.f. cavities (blue), 
and the trajectory of the reference particle (red). }
\label{fig:unitside}
\end{figure}

The double flip cooling channel is a system consisting of three 
homogeneous solenoids with two field-flip sections.
The first flip occurs at a 
relatively small magnetic field, B=3~T, to keep the longitudinal motion 
under control. The field is then increased adiabatically from -3 to -7~T, and 
a second field flip performed at B=7~T. 
Figure~\ref{fig:unitside} shows a side view of a lattice unit cell, 
consisting of a six 201~MHz Pill Box cavities linac and one liquid hydrogen
absorber, inside a solenoid. 
Details on the design and performance of this channel are 
available in Ref.~\cite{pacDF}.

\subsection{The Helical Channel}

\begin{figure}[h]
\vskip-9.3cm
\vspace{1mm}
\hskip4cm
\epsfxsize=2.6in
\centerline{\epsfbox{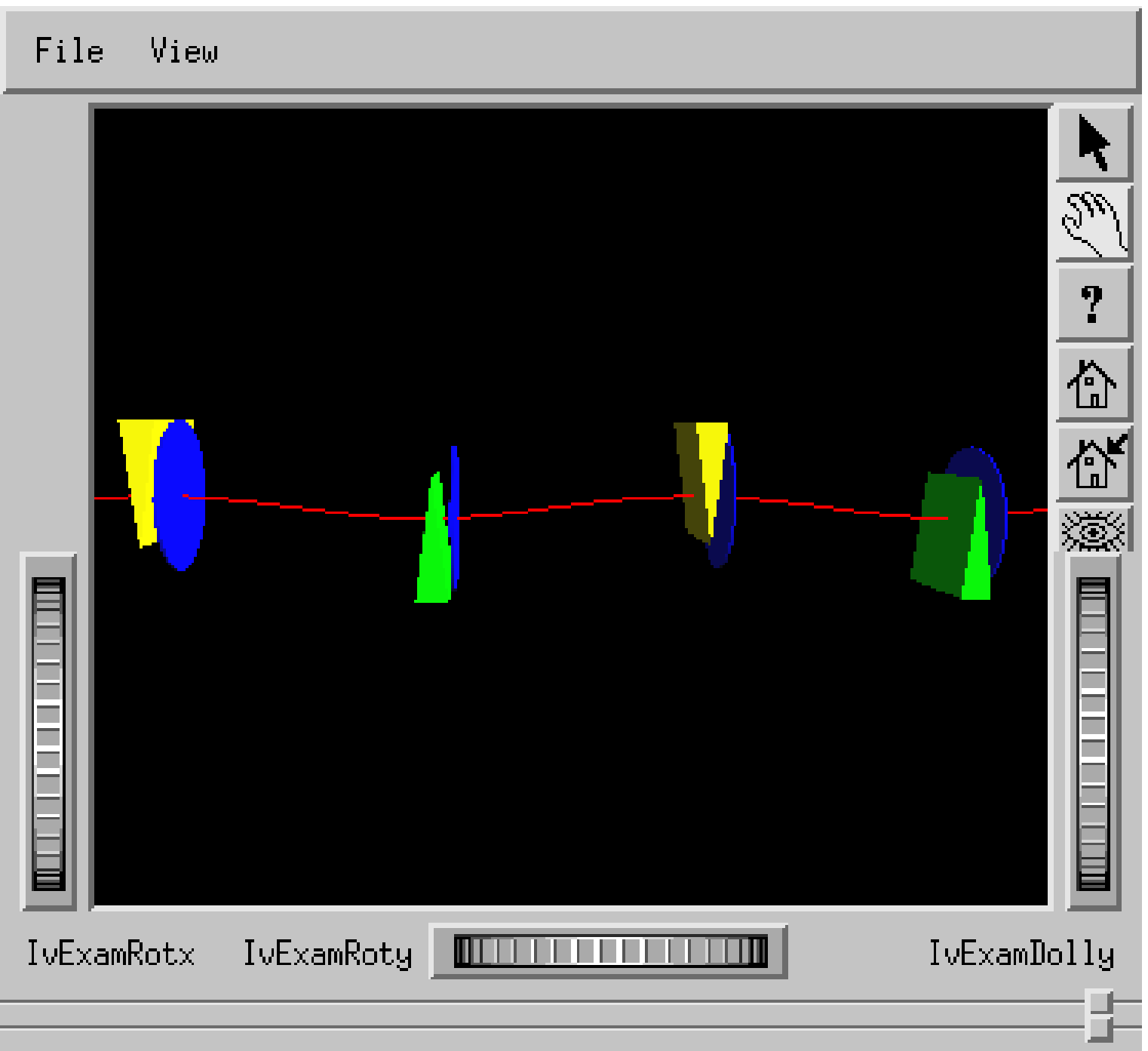}}
\vskip2.2cm
\label{fig:heliside}
\end{figure}

The helical channel cools both in the transverse and longitudinal
directions. The lattice is based on a long solenoid with the 
addition of a rotating transverse dipole field, lithium hydride wedge 
absorbers, and 201~MHz r.f. cavities. 
Figure~\ref{fig:heliside} shows a side view of the helical channel, including
the wedge absorbers, idealistic (thin) r.f. cavities, and the trajectory
of the reference particle. 
The design details and performance of this channel are 
described in Ref.~\cite{pacHE}.

\subsection{The Low Frequency Channel}

This is a design based on 44/88 MHz r.f. technology. 
A unit cell is composed of four solenoids embedded in  four r.f. cavities, 
followed by a liquid hydrogen absorber. 
Figure~\ref{fig:berywin} shows a unit cell of the low frequency channel,
including the solenoids, the absorber, and the relevant section of
the r.f. field map (inside the magnets). More information about
this channel may be found in Ref.~\cite{mc230}.

\subsection{Other Systems}

Among other simulations performed with the Beam Tools for Geant4 we may cite:
the {\it Alternate Solenoid Channel (sFoFo)}~\cite{pacSF}, and a {\it High
Frequency Buncher/Phase Rotator} scheme for the neutrino 
factory~\cite{mc253,mc254}.

\section{Summary}

The Beam Physics Tools for Geant4 are used in numerous accelerator studies,
reported in conference proceedings and proposals. Geant4 is especially suited
to systems where accelerators, shielding, and detectors must be studied jointly
with a simulation.
The Beam Tool libraries, a software reference manual, and a user's guide,
are available from the Fermilab Geant4 web page~\cite{fermiG4}.

\ack
We thank  Mark Fishler, for contributing the data cards and
and spline fit classes, Jeff Kallenbach for helping with visualization
issues, and Walter Brown for providing C++ consultancy. We are also 
grateful to the Geant4 Collaboration for answering our questions. 
In particular, we thank J.~Apostolakis, M.~Asai, G.~Cosmo, M.~Maire, L.~Urban, 
and V.~Grichine.

\end{document}